\documentclass[prl,aps,twocolumn]{revtex4}
\usepackage{graphicx}
\begin{document}
\title{Quantum Interference Noise Near the Dirac Point in Graphene}
\author{Atikur Rahman}
\author{Janice Wynn Guikema}
\author{Nina Markovi\'c}
\affiliation{{\it Department of Physics and Astronomy, Johns Hopkins
University, Baltimore, Maryland 21218, USA.}}
\begin{abstract}
Effects of disorder on the electronic transport properties of graphene are
strongly affected by the Dirac nature of the charge carriers in graphene. This is
particularly pronounced near the Dirac
point, where relativistic charge carriers cannot efficiently screen the impurity potential.
We have studied time-dependent
conductance fluctuations and magnetoresistance in graphene in the close vicinity
of the Dirac point. We show that the fluctuations are due to the quantum interference
effects due to scattering on impurities, and find an unusually large reduction of the relative noise power in magnetic
field, possibly indicating that an additional symmetry plays an important role in this regime.

\end{abstract}
\pacs{73.20.Mf, 73.20.Qt, 73.61.Ph, 73.63.-b}
\maketitle

In disordered electronic systems, quantum corrections to the conductance arise
due to quantum interference between paths of electrons scattered on
random impurities. In the absence of a magnetic field, the
electron paths that traverse the loops in a clockwise fashion
interfere constructively with the counterclockwise paths through the
same loops, resulting in a small change in the conductance.
Specifically, the backscattered paths (the paths that return to
their origin) lead to a correction to the average conductance of the
system, known as weak localization (WL) \cite{1,2,3}. Magnetic field
adds a different phase factor to the paths that are identical, but
traversed in the opposite sense, removing the WL corrections, but
one still observes the universal conductance fluctuations (UCF) as a
function of magnetic field or chemical potential, which arise from
adding the interference contributions from all possible paths
\cite{4,5}. The quantum interference contribution to the conductance
also fluctuates if the impurity configuration changes over {\it
time}, leading to time-dependent conductance fluctuations that are
expected to cause $1/f$ noise \cite{6,7,8}.

In graphene, the quantum interference phenomena are affected by the
pseudospin and valley degrees of freedom \cite{9,10}. Conservation
of pseudospin precludes backscattering, suppressing WL and causing
weak antilocalization (WAL) \cite{11}, while intervalley
scattering restores the WL \cite{12,13,14}. Additional
effects, such as defects and corrugations, can
completely suppress the quantum corrections \cite{15}. Depending on
the carrier density and the nature of the disorder, all three
regimes (WL, WAL and the suppression of quantum corrections) are
observed experimentally \cite{16,17}.
UCF in graphene depend on the carrier density and the nature of the impurity scattering,
but can also depend on the details and the geometry of the sample \cite{Fal'ko,18,Horsell}. In particular,
strong intervalley scattering is found to suppress UCF \cite{Fal'ko, 18},
in contrast to its effect on WL. However, the majority of the theoretical and experimental work on quantum
corrections has focused on the regime away from the Dirac point. In the close vicinity to the Dirac point
(at low doping and low temperatures), the relativistic Dirac quasiparticles are unable to screen the long-range
Coulomb interactions in the usual way, altering electron-electron interactions \cite{Kotov}.

In this work, we describe measurements of the time-dependent conductance fluctuations in graphene, focusing 
specifically on
the low-carrier density regime near the Dirac point.  We find that the $1/f$ noise is reduced in magnetic
field, with a characteristic field and temperature dependence that suggests quantum interference as the origin of the noise. However, the observed relative noise reduction is twice as large as what one
might expect based on the fundamental symmetry considerations and the current theoretical understanding of quantum 
transport in graphene.

\begin{figure}
  \includegraphics[width=8 cm]{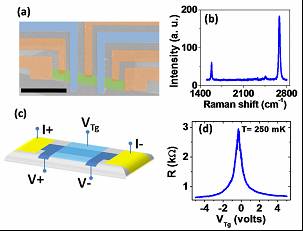}
  \caption{{\bf (a)} False color scanning electron microscope image
  of a typical single layer device (SL1). The graphene flake is highlighted
  in green and the top gates are shown in blue. The distance between
  the voltage leads is typically around one micrometer. The scale
  bar is 3 $\mu$m long.
  {\bf(b)} Raman spectra of one of the samples. The observed peak structure
  is characteristic for single layer graphene. {\bf(c)} Schematic of the
  four probe measurement setup with external voltage probes.
  {\bf(d)} Resistance as a function of top gate voltage for zero back
  gate voltage for sample SL1. The Dirac point, or the charge
  neutrality point, is located at $V_{Tg}=-0.3$ V.}
  \label{0}
\end{figure}

A scanning electron microscope image of a typical top gated device
(SL1) is shown in Fig.\ 1(a) (see Supporting Information for fabrication details).
Raman spectroscopy was used to confirm that all samples were single
layer graphene, with typical results shown in Fig.\ 1(b). Electrical
measurements were done in a 4-probe geometry, using external voltage
probes \cite{24} as shown on the schematic in Fig.\ 1(c). The
typical resistance as a function of top gate voltage ($V_{Tg}$) with
zero back gate voltage ($V_{Bg}$) is shown in  Fig.\ 1(d). The peak
in the resistance occurs at the Dirac point, at the value of top
gate voltage -0.3 V, at which the carrier density in graphene
reaches the minimum.

\begin{figure}
    \includegraphics[width=8 cm]{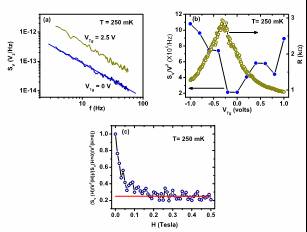}
  \caption{{\bf(a)} Noise power as a function of frequency
  (plotted on a log-log scale) for two different values
  of the top gate voltage. Straight line shows the 1/f dependence
  of the noise spectra. {\bf(b)} Noise power (left) and resistance
  (right) as a function of top gate voltage in the vicinity of
  the Dirac point.
  {\bf(c)} Normalized noise power as a function of magnetic field for
  a single layer graphene device. It is evident that the relative noise is reduced by a factor
  of four from its zero magnetic field value above a certain field.
  The straight line indicates the reduction of the zero-field value by a factor of four.}
  \label{2}
\end{figure}

Low-frequency noise measurements were done using the ac noise
measurement technique \cite{25}. The measured noise power ($S_V$)
showed $1/f^{\alpha}$ dependence for either gate (top or back)
voltage with values of $\alpha$ close to 1 (Fig.\ 2(a)). We found
that the noise data were highly reproducible over time at any
temperature and did not depend on the direction or scan step of the
gate voltage. The normalized noise power density ($= fS_V/V^2$ or
$fS_I/I^2$) was found to be independent of the bias current or
voltage (see Supporting information, Fig.\ S1), ruling out any
issues due to heating by the bias current.

The normalized noise power as a function of the top gate voltage in
the vicinity of the Dirac point is shown in Fig.\ 2(b). We find that
the noise decreases upon approaching the Dirac point from both
directions, reaching a minimum close to the Dirac point.

When a magnetic field is applied perpendicular to the substrate, the
noise power decreases rapidly as shown in Fig.\ 2(c). After reaching
some characteristic value of the field, the relative noise power
saturates at the value that is a factor of four smaller than the
zero-field value. Assuming that the characteristic field corresponds
to threading one flux quantum through a phase-coherent area of the
sample, we find the phase coherence length ($\L_{\phi}$) to be in the
range between $200-300\; nm$ for our samples.

\begin{figure}
  \includegraphics[width=8 cm]{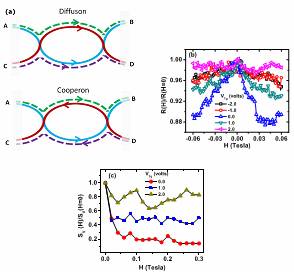}
  \caption{{\bf(a)} Schematic of pairs of electron trajectories
  that form closed loops. The conductance fluctuations are
  caused by interference between paths from A to B and those
  from C to D that intersect somewhere in the interior of the
  sample. The Diffuson contribution is shown in the upper
  panel: all the paths in the loop are traversed in the same
  sense, and the magnetic field does not introduce a relative
  phase factor. In the Cooperon channel, shown in the bottom
  panel, the magnetic field introduces a relative phase when
  the loop is traversed in the opposite sense. In this case,
  contributions from various loops no longer add in a coherent
  way and the quantum corrections to conductivity disappear.
  {\bf(b)} Magnetoresistance as a function of magnetic field for
  various top gate voltages with zero back gate voltage (sample SL1).
  A negative magnetoresistance is observed in the vicinity
  of the Dirac point, with the maximum of magnetoresistance
  observed close to zero top gate.  As the gate voltage is
  increased in both directions, the magnetoresistance decreases.
  {\bf(c)} Relative noise power for different top gate voltages as a
  function of a magnetic field (sample SL1). The reduction of the relative
  noise power by about a factor of four is observed at zero top gate,
  but this factor decreases for larger values of top gate voltage. }
  \label{6}
\end{figure}

The low-frequency noise was studied experimentally in single-layer
graphene transistors and was suggested to be related to the
fluctuating charges in the vicinity of graphene \cite{21,22,23}.
However, the reduction of the noise power upon application of a
small magnetic field observed here is a strong indication that
quantum interference effects dominate the low-frequency noise. In
the case of disordered metals, the conductance fluctuations can be
calculated by considering all possible trajectories that an electron
can take while scattering off of random impurities. Particularly
important are the combinations of paths that connect two different
points in a sample, as shown in Fig.\ 3(a).  There are two
contributions to the interference between paths from A to B and
those from C to D: diffuson and cooperon contribution. The diffuson
contribution is insensitive to the magnetic field, as no relative
phase is introduced between the various paths. On the contrary, the
magnetic field removes the cooperon contribution, reducing the number of
conduction channels by a factor of two and reducing the
relative noise by precisely a factor of two \cite{8}.

In the same regime, we find negative magnetoresistance as a function of magnetic field 
at different gate voltages, is shown in
Fig.\ 3(b). Negative magnetoresistance may be due to WL, but we
observe it only in the narrow range of gate voltages in the vicinity
of the Dirac point - the negative magnetoresistance decreases as the
gate voltage is increased in both directions. The effect of magnetic
field on noise also depends on the gate voltage as shown in Fig.\
3(c).  Away from the Dirac point, the noise becomes less sensitive
to the magnetic field for both positive and negative gate voltages.
The noise characteristics are found to be symmetric with respect to
magnetic field, and similar results were found for both back gate
and top gate. The four-fold reduction in the noise is not always
observed precisely at the Dirac point, but it coincides exactly with the gate
voltage at which the maximum negative magnetoresistance is also observed 
(at small positive gate voltages relative to the Dirac point).
Similar behavior is observed as a function of back gate voltage (see
Supporting information, Fig.\ S2). The overall change in resistance
is small compared to the change in the noise (see Supporting
information, Fig.\ S3).

According to the present understanding, WL can be observed in
graphene in the presence of strong intervalley scattering, which can
arise due to atomically sharp potentials (such as edges or atomic defects).
In the case of strong intervalley scattering, many aspects of
quantum transport in graphene are expected to be identical to those
in disordered metals \cite{10,14,Fal'ko, 18}. In particular, the conductance fluctuations
should exhibit universal properties, as they depend only
on the symmetries of the random ensembles that describe the
disordered system, and not on their detailed configuration.
The variance of the interference-induced conductance fluctuations in graphene
will generally have a prefactor that depends on the interplay of inelastic and elastic scattering
lengths and the shape of the sample \cite{Fal'ko, 18}, but
graphene with broken valley symmetry should belong to
the orthogonal Wigner-Dyson symmetry class in the absence of a
magnetic field \cite{10}. Application of a magnetic field will put
it in the unitary symmetry class, and a two-fold reduction in the
relative noise power will be expected on general symmetry
grounds.

Additional two-fold reduction in the relative noise power is
expected when the Zeeman energy exceeds $hD/{L_\phi}^2$ \cite{8}.
The two-fold reduction in the noise power has been observed in
metals \cite{19,20}, as well as the additional two-fold reduction at
a larger magnetic field due to Zeeman splitting \cite{26,27}. In our
samples, the Zeeman splitting cannot explain the four-fold
reduction, which is observed for small characteristic fields (50
mT), where the Zeeman splitting (0.006 meV) is smaller than both the
thermal energy (0.02 meV) and $hD/{L_\phi}^2$ (0.08 meV).

\begin{figure}
  \includegraphics[width=8cm]{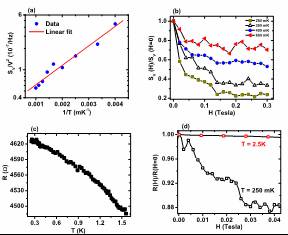}
  \caption{{\bf(a)} Normalized noise power is shown as a
  function of inverse temperature on a log-linear scale.
  The line represents a linear fit, and the normalized
  noise power shows a $\exp(1/T)$ dependence on temperature.
  {\bf(b)} Relative noise power is plotted as a function of magnetic
  field for different temperatures. The reduction of the relative
  noise power by a factor of four is observed at 250mK, but the
  reduction is smaller at higher temperatures. {\bf(c)} Resistance as a
  function of temperature is shown for $V_{Tg}, V_{Bg}=0$, in the
  regime highlighted in Fig.\ 2 (b). The slight increase of the
  resistance with decreasing temperature indicates insulating behavior.
  {\bf(d)} Magnetoresistance as a function of magnetic field is shown for
  two different temperatures. It is evident that a much smaller
  magnetoresistance is observed at higher temperatures.}
  \label{3}
\end{figure}

The temperature dependence of the noise is also unusual in this
regime. For normal metals in the phase coherent regime, the noise
due to fluctuating scatterers depends on temperature as $T^{-1}$, as
observed in several systems \cite{19,20}. We found that the noise
decreases with increasing temperature and the normalized noise power
shows a $\exp(1/T)$ dependence, as shown in Fig.\ 4(a) (similar
dependence was also found in other work \cite{28}). As the
temperature increases, the relative noise power is still reduced in
magnetic field, but by a smaller factor, as shown in Fig.\ 4(b). The
sample resistance increases slightly with decreasing temperature, as
shown in Fig.\ 4(c), so the temperature dependence of the noise
cannot be explained by the resistance change. The slowly increasing
resistance with decreasing temperature is consistent with WL,
as is the fact that the negative magnetoresistance also decreases
with increasing temperature, as shown in Fig.\ 4(d).

It is well known that charge-inhomogeneous regions (puddles) tend to form
in the vicinity of the Dirac point \cite{Martin}. The presence of the top gate
also locally dopes the graphene, forming pn junctions at the edges. A random
network of puddles or pn junctions could be expected to show conductance fluctuations
and magnetoresistance that reflect the fluctuations in the electrostatic
environment \cite{Cheianov1, Cheianov2}. However, one might expect such fluctuations
to increase with temperature, leading to the increase of the noise power with increasing temperature,
which is in contradiction to our observations. The temperature dependence of the resistance,
magnetoresistance and the relative noise power reduction in magnetic field are all consistent with
a decrease of the phase coherence length as the temperature is increased. In addition,
the observation of the Aharonov-Bohm oscillations in similar samples confirms that both
the electron and the hole transport is phase-coherent across pn junctions and any electron-hole
puddles in the vicinity of the Dirac point \cite{Rahman}.

The decrease of the quantum interference noise by a factor of four
in magnetic field is not presently understood, but the unusual nature of
quantum corrections near the Dirac point may offer insight into phenomena
observed in other experiments, such as the four-fold decrease of mobility depending on 
the nature of impurity scattering \cite{29}, or
the anomalous backscattering \cite{30, 31} near the Dirac point. A better understanding of 
the quantum interference 
noise may also provide
useful clues about the nature of the impurity scattering in this regime.

\vskip 0.2in

The authors would like to thank A. Morpurgo, E. McCann, F. Guinea,
V. Fal'ko and I. Aleiner for useful comments and suggestions. N. M.
would like to thank the Aspen Center for Physics and the NSF Grant 1066293. J. W. G. was
supported in part by the M. Hildred Blewett Fellowship of the
American Physical Society.

\end{document}